\documentclass[12pt]{article}

\usepackage{amsfonts,amsmath}
\usepackage[mathscr]{eucal}
\usepackage{amssymb}
\usepackage{amsthm}
\usepackage{epsfig,epsf}
\usepackage{graphicx}
\usepackage{graphics}
\usepackage{bbold}
\theoremstyle{plain}

\newcommand{\be}{\begin{eqnarray}}
\newcommand{\ee}{\end{eqnarray}}
\newcommand{\nn}{\nonumber \\}
\newcommand{\lb}{\label}
\newcommand{\p}[1]{(\ref{#1})}
\newcommand{\vecg}[1]{\mbox{\boldmath $#1$}}

\newcommand\cL{{\cal L}}
\newcommand\cN{{\cal N}}

\begin{document}
\begin{titlepage}

\renewcommand{\thefootnote}{\star}
\begin{center}

\vspace{0.1cm}

{\LARGE\bf  Ultraviolet divergences in non-renormalizable  supersymmetric theories}

\vspace{1.2cm}
\renewcommand{\thefootnote}{$\star$}

{\large\bf Andrei~Smilga} \vspace{0.3cm}

{\it SUBATECH, Universit\'e de Nantes,}\\
{\it 4 rue Alfred Kastler, BP 20722, Nantes 44307, France;}\\
\vspace{0.1cm}

{\tt smilga@subatech.in2p3.fr}\\

\end{center}

\vspace{0.2cm} \vskip 0.6truecm \nopagebreak

\begin{abstract}
\noindent

We present a pedagogical review of our current understanding of the ultraviolet
structure of ${\cal N} = (1,1)$ 6D supersymmetric Yang-Mills theory and of
${\cal N} = 8$ $4D$ supergravity. These theories 
are not renormalizable, they involve
 power ultraviolet divergences and, in all probability, an 
infinite set of higher-dimensional counterterms that contribute to on-mass-shell scattering
amplitudes. 

A specific feature of
 supersymmetric theories (especially, of extended supersymmetric theories) 
is that these counterterms may not be invariant
off shell under the full set of supersymmetry transformations. 
The lowest-dimensional nontrivial 
counterterm is supersymmetric on shell. Still higher counterterms may 
lose even the 
on-shell invariance. On the 
other hand, the full effective Lagrangian, generating the amplitudes and 
representing an infinite sum of counterterms, still 
enjoys the complete symmetry of original theory.

 We also discuss simple supersymmetric quantum-mechanical models that exhibit the same behaviour.

\end{abstract}
\vspace{0.3cm}

\newpage

\end{titlepage}

\setcounter{footnote}{0}

\setcounter{equation}0

\section{Introduction}

The standard Einstein gravity and its supersymmetric extensions involve 
a dimensionful constant and are not renormalizable theories. Still, several years ago,
 certain hopes were expressed in the conference talks and in the literature that the
extended ${\cal N} = 8$ Poincar\'e supergravity might be a ``finite theory'', meaning that
it does not involve relevant higher-dimensional counterterms.

To a considerable extent, these hopes 
were based on the explicit calculations \cite{Bern-grav}, which demonstrated the 
absence of logarithmic divergences in on-mass-shell scattering amplitudes through 4 loops.
This remarkable cancellation is explained by very high symmetry of the theory. This symmetry
simply does not allow one to write down the counterterms of dimension\footnote{The
 word {\it dimension} will be used in this paper in two meaning. First, it is space-time
 dimension which is denoted by the
capital $D$. Second, it is dimension of various operators which we will denote by the low-case $d$.}
$d = 4,6,8,10,12,14$ which enjoy on shell the general covariance, the extended ${\cal N} = 8$ 
supersymmetry and different dualities, and which could give rise to divergences.

However, most experts believe nowadays that, at the level $d=16$ (corresponding to 7 loops) or, 
at any rate, at the level $d=18$ (corresponding to 8 loops), such counterterms do appear and generate 
divergences.~\footnote{We will discuss it later, but we
hasten to say right now that all the nontrivialities mentioned above
concern the {\it logarithmic} divergences. A theory with a dimensionful constant necessarily has 
{\it power} ultraviolet divergences, which are not associated to higher-dimensional counterterms. 
The calculations in \cite{Bern-grav} are not sensitive to power divergences.}

${\cal N} = 8$ supergravity is a very complicated theory. That is why it is interesting to study
a much simpler model, the  ${\cal N} = (1,1)$ 6D supersymmetric Yang-Mills (SYM) theory. Its coupling
constant also carries dimension, so that the theory is not renormalizable. Nontrivial 
counterterms generating logarithmic divergences in the amplitude already appear 
there at the 3-loop level. 
The presence of these divergences was confirmed by explicit perturbative calculations in 
\cite{Bern-gauge}.

As was already mentioned in the Abstract, non-renormalizable theories with extended 
supersymmetry differ from the well-known non-renormalizable non-supersymmetric theories 
(Einstein's gravity, 
Fermi theory of weak interactions, chiral theory describing low-energy pion physics) by the fact
that extended supersymmetry cannot be kept {\it off shell} for higher-dimensional counterterms.
Only the full effective Lagrangian, representing an infinite sum of counterterms of higher and higher
dimension, keeps this symmetry.
This fact is known to experts, but there is no clearly written and not too technical text where
it would be pedagogically explained. That is what we will try to do in this 
lecture.~\footnote{ A reader interested in more technical issues should consult 
the recent paper \cite{BIS}, the logics of which we mostly follow here.}

We start in Sect.~2 
by saying few words about generic features of non-supersymmetric
non-renormalizable theories: chiral theory, Fermi theory  and Einstein's gravity. 
 The main observation:
one can in principle carry out the renormalization program and get rid of 
ultraviolet divergences also in a non-renormalizable theory, redefining 
order by order an infinite set of couplings, but it does not help one to
 calculate scattering amplitudes at the energies exceeding the {\it unitary limit} --- 
an intrinsic feature of all non-renormalizable theories. 

A general wisdom is that complicated field-theoretical phenomena can be 
much better understood by 
studying  toy quantum-mechanical models having similar features. 
 In Sect.~3, we consider the simplest supersymmetric quantum-mechanical (SQM) 
model, involving only one real supervariable
  \be
\lb{SQM-X}
 X \ =\ x + \theta \bar \psi + \psi \bar \theta + F \theta \bar \theta \, .
  \ee

We note in particular that the auxiliary field $F$ can be algebraically 
excluded only in the simplest Witten's version of this model with the bosonic 
kinetic term $\propto \dot{x}^2$ \cite{Witten81}.  If the Lagrangian involves 
higher derivatives, the field $F$ becomes dynamical. 

Generically, one cannot get rid of $F$ in this case. However, if 
the action represents the sum of the Witten term and a higher derivative term,
the field $F$ can be formally excluded via an infinite perturbative series. 
The Lagrangian thus obtained involves only the fields $x, \psi,  \bar\psi$ and
their time derivatives of all orders. It is invariant under {\it modified} 
(compared to Witten's model) supersymmetry transformations that also include time
derivatives of all orders.
We discuss then the implications of this simple observation for ${\cal N} =1$,
 ${\cal N} =2$ and  ${\cal N} =4$ $4D$ SYM theories.

Sect.~4 is devoted to other instructive  models. First, we discuss the so-called 
{\it maximally supersymmetric} SQM model. It is obtained by dimensional reduction
from  ${\cal N} =1$ $10D$ SYM theory and has 16 real supercharges.
The model involves Abelian flat directions or {\it vacuum valleys}, where the 
non-Abelian field strength and the associated potential vanish. In the bottom
of this valley, far enough from the origin, the dynamics of the model is 
described by the effective low-energy Born-Oppenheimer (BO) Hamiltonian, 
representing an infinite series over the small BO parameter 
$\gamma$. The same is true for the corresponding Lagrangian,~\footnote{We will see in Sect.~3 that, with
a natural definition of $\gamma$, 
the series goes over $\gamma^{3n}$.} 
 \be
\lb{L-series}
 L^{\rm eff} \ =\ L_0 + \gamma^3 L_1 + \gamma^6 L_2 + \ldots 
 \ee
  The leading term has a simple form and is invariant (up to a time derivative)
under supersymmetry 
transformations that also have a simple form. The whole series is 
invariant under the modified transformations, which also represent an 
infinite series in $\gamma$,
  \be
\lb{delta-series}
\delta \ =\ \delta_0 + \gamma^3 \delta_1 + \gamma^6 \delta_2 + \ldots \, .
  \ee
 But an individual term $L_{n\geq 1}$ in 
\p{L-series} is {\it not} supersymmetric. On the other hand, 
one can observe that
$L_1$ is invariant under the same supersymmetry transformations as $L_0$ if
the dynamical variables satisfy the equations of motion following from $L_0$.
This is the SQM counterpart of on-shell supersymmetry of the counterterms
in $6D$ SYM and $4D$ supergravity.

At the end of Sect.~4, we go back to non-supersymmetric chiral field theory and study the structure of its 
{\it tree}
amplitudes. We consider the 
 leading term \p{Lchiral-tree} of the chiral Lagrangian and observe that, in close analogy with \p{L-series}, 
it can be represented as an 
infinite series of the
terms of growing dimension, such that an individual term in this series does not have the full 
$SU_L(2) \times SU_R(2)$ symmetry of \p{Lchiral-tree}. The next-to leading term in this expansion enjoys,
however, the full chiral symmetry {\it on shell} --- in exactly the same way as the term $L_1$ in the
series \p{L-series} enjoys on shell the full extended supersymmetry.

Note that still higher terms $L_{2,3,\ldots}$ in this expansion need not to be on-shell symmetric.
The scattering amplitudes keep, however, complete chiral symmetry at all orders.

 In Sect.~5, we discuss $6D$ SYM theories, both $\cN = (1,0)$ and $\cN = (1,1)$ versions thereof.
The $\cN= (1,0)$ theory involves a chiral anomaly that breaks gauge invariance. The $\cN = (1,1)$ theory
is anomaly-free, but is not renormalizable because of dimensionful coupling. 
 We briefly describe the harmonic superfield formalism, allowing one to understand the symmetry structure
of these theories in the most clear way and to write down explicit closed expressions for the actions.

In Sect.~6, we discuss the structure of higher-dimensional  counterterms. 
The relevant 
counterterms appear at the 3-loop level, having canonical dimension $d=10$.
 They are invariant under $\cN = (1,1)$ supersymmetry 
transformations only on shell, but not off shell. One can write two different such counterterms.
Surprisingly, the presence of only one of them was ``observed'' in explicit loop calculations. The reason
for such an unexpected cancellation is not clear yet. 

Finally, in Sect.~7 we briefly discuss the situation in supergravity. 
Logarithmic UV divergences  and associated counterterms probably
appear there at the 7-loop or 8-loop level. 
In supergravity, total cross sections diverge, and it is difficult to define
simple observables whose energy dependence could be studied.
 But the presence of 
higher-dimensional counterterms will definitely prohibit crossing
 the Planck mass barrier and performing meaningful perturbative calculations 
 at high energies.

\section{Non-supersymmetric theories}
\setcounter{equation}0

Historically, the first non-renormalizable field theory model that attracted the attention of theorists
was Fermi's 4-fermion model characterized by the dimensionful constant $G_F$. But we choose to 
 discuss in some
more detail the effective chiral theory developped in Ref. \cite{chiral} and used thereafter for many practical
calculations in  low-energy QCD.
  
To leading order and neglecting $u$ and $d$ quark  masses (so that pions are also massless), the effective chiral
Lagrangian describing pion interactions reads
  \be
\lb{Lchiral-tree}
  {\cal L}^{(0)} \ =\ \frac {F_\pi^2}4  \,  \langle \partial_\mu U \partial^\mu U^\dagger \rangle \, ,
  \ee
where 
 \be 
\lb{U-SU2}
U(x) = \exp \left\{
\frac {i \pi^a(x) \sigma^a}{ F_\pi} \right \} 
 \ee
is an $SU(2)$ matrix,  $\pi^{a=1,2,3}(x)$ are the pion fields, 
$F_\pi = 93$ MeV is the pion decay constant and $\langle \cdots \rangle$ stands for the  trace. 

The Lagrangian possesses $SU_L(2) \times SU_R(2)$ symmetry --- it is invariant under a 
multiplication of $U$ on an arbitrary
unitary matrix on the left or on the right.~\footnote{There is also a $SU_L(3) \times SU_R(3)$ version
of Eq.\p{Lchiral-tree} describing interactions of the mesons 
of the pseudoscalar octet, but we do not need to discuss it.}
 Expanding the exponential, one can derive,
  \be
\lb{Lchiral-expan}
 {\cal L}^{(0)}  \ =\ \frac 12 (\partial_\mu \pi^a)^2 + \frac 1{6 F_\pi^2}
 \left[ (\pi^a \partial_\mu \pi^a)^2 - (\pi^a \pi^a)
(\partial_\mu \pi^b)^2 \right] + \cdots
  \ee
The quartic term in the Lagrangian involves derivatives, and the 
$\pi \pi$ scattering amplitude grows with energy,
 \be
\lb{Mtree-pion}
M^{(0)}_{\pi\pi \to \pi\pi} \ \sim \ \frac{ E^2}{F_\pi^2} \, .
  \ee
  This model is not 
renormalizable and involves power divergences, which show up in the loops. 

People are  often not concerned about power divergences, because the latter 
do not arise when the dimensional regularization (technically,
 the most simple method to calculate the Feynman graphs) is used.
 But to disregard 
them completely
 amounts to hiding the problem under the carpet. The best and the most physical,
to our mind, regularization scheme is the lattice regularization. Anyway, to 
attribute a meaning to the path integral symbol, one should discretize 
the space-time and define the path integral as a 
continuum limit of a finite-dimensional integral.~\footnote{This works for
conventional field theories in flat space-time. How to define path integral for
gravity is a separate difficult question that we are not going to address here.}
 Using lattice regularization for non-renormalizable theories exhibits power 
divergences \cite{Shush}. The same is true for Slavnov's higher-derivative
regularization scheme used usually for gauge theories \cite{Slavnov}. 
Power divergences  also appear for 
renormalizable theories including interacting scalar fields: recall the fine
tuning problem in the Standard Model and the hierarchy problem in
 non-supersymmetric models of Grand Unification.

On top of power divergences, there might also  be logarithmic
divergences. In chiral theory, they already show up at the 1-loop level. 
The one-loop contribution to the amplitude
reads
   \be
\lb{scat-pion}
 M^{(1)}_{\pi\pi\to \pi\pi}  \ =\ \frac {\Lambda^2}{F_\pi^2} M_{\pi\pi\to \pi\pi}^{(0)}
\, + \, 
\frac {A(s,t)} {F_\pi^4} \ln \frac \Lambda \mu  \, +  \, \frac {B(s,t,\mu)}{F_\pi^4} 
   \ee 
with an arbitrarily chosen $\mu$, on which  $B$ also 
 logarithmically depends.

\begin{figure} [t]
\begin{center}
\includegraphics[width=5in]{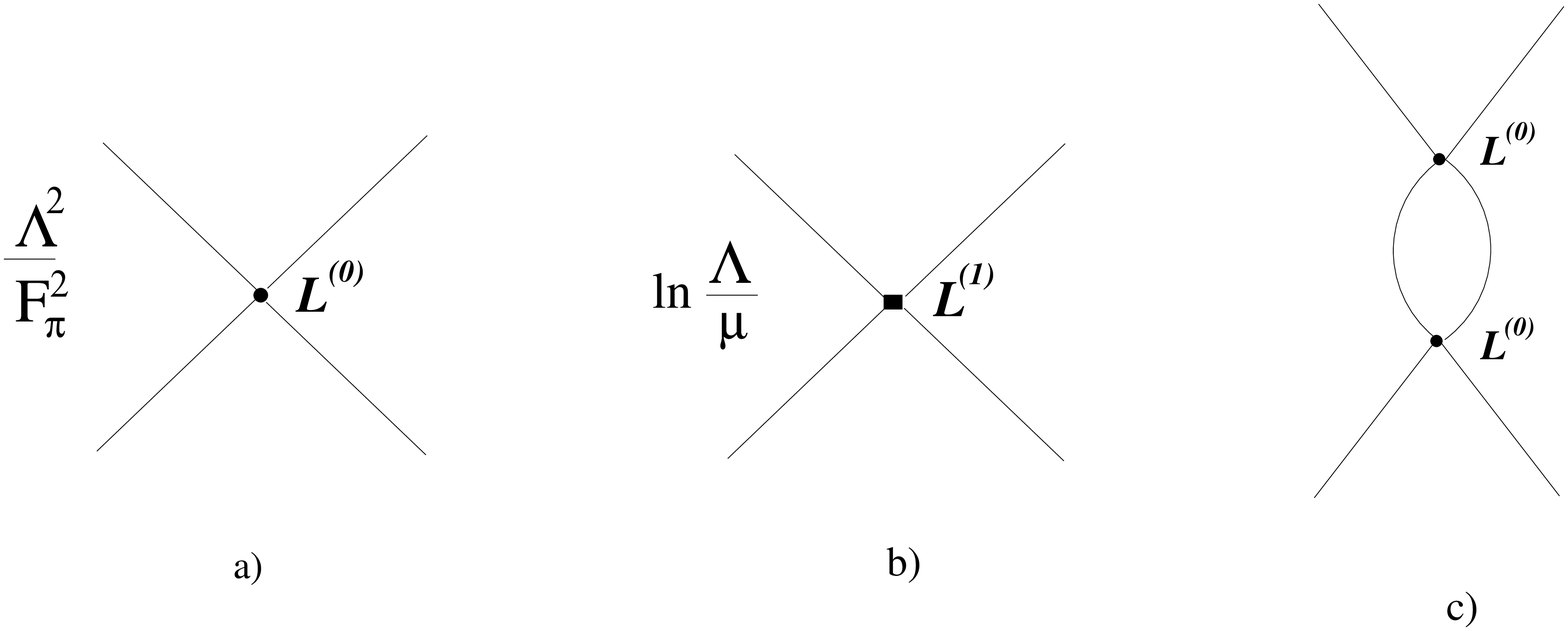}
\end{center}
\caption{One-loop contributions to the pion scattering amplitudes. {\it a)}
renormalization of $F_\pi^{-2}$, {\it b)} logarithmically divergent 
local contributions, {\it c)} UV-finite nonlocal contributions.}
\label{chiral-1loop}
\end{figure}

The last UV-finite
term in \p{scat-pion} is nonlocal and complicated. The first two terms are local, however.
 Indeed, to obtain an ultraviolet 
divergence, the characteric loop momenta should be large, and then a complicated loop 
graph is effectively shrinked to a point.
We arrive at the notion of the {\it Wilsonian effective Lagrangian} that generates local 
contributions to the scattering amplitudes.  The amplitude $ M^{(0)}_{\pi\pi\to \pi\pi}  $ 
is generated by \p{Lchiral-tree} (and, at the 1-loop level, the coupling constant $F_\pi^2$ is renormalized, including power divergences), while
 the contribution $\sim A(s,t)$ 
 is generated by two different higher-dimensional {\it counterterms}, 
 \be
\lb{Lchiral-1loop}
  {\cal L}^{(1)}_1 \ =\ \langle \partial^\mu U^\dagger \partial_\mu U \rangle ^2\, ,
\ \ \ \ \ \ \ \ \ \ {\cal L}^{(1)}_2 \ =\ \langle \partial_\mu U^\dagger \partial_\nu U \rangle 
 \langle \partial^\mu U^\dagger \partial^\nu U \rangle \, . 
\ee
 Different contributions to the amplitude are schematically represented in Fig.~\ref{chiral-1loop}.

Note that one could, in principle, write down several other counterterms 
involving four derivatives. But these extra terms vanish when the field $U$ satisfies the tree equations of motion,
  \be
\lb{eqmot-chiral}
 (\Box U) U^\dagger - U (\Box U^\dagger) \ =\ 0 \, ,
 \ee
As a result, these extra counterterms do not contribute to the {\it on-shell} scattering amplitudes and are irrelevant.~\footnote{One can always get rid of the counterterms proportional to the tree equations of motion by redefining the fields. Schematically,
  \be
{\cal L}(\Phi) + \frac {\delta {\cal L}} {\delta \Phi} A(\Phi)  \ \approx \   {\cal L}(\Phi + A(\Phi) )  \, .
  \ee
}

Both in renormalizable and non-renormalizable theories, one can in principle get rid of all UV
divergences, if hiding them in the constants of the bare Lagrangian, so that the renormalized constants would 
be UV-finite. For a non-renormalizable theory, one should do it for an infinite number of the bare
constants associated with the infinite number of relevant counterterms. 
Even though this procedure looks rather unnatural, it
allows one to treat a non-renormalizable theory perturbatively, and not
 only at the tree level, but also with
taking loops into account.

It is well-known, however, that such calculations make sense only at 
low enough energies. In non-renormalizable theories, tree
amplitude rapidly grow with energy. For example, the tree pion
 scattering amplitude behaves as in \p{Mtree-pion}.
But this growth cannot keep going  indefinitely. This would violate unitarity --- cross sections
cannot grow faster than $\propto \ln^2 E$.~\footnote{One can prove it
if the spectrum does not involve massless particles  \cite{Froissart}. To apply this argument
for the chiral theory, one should endow the pions with the mass by adding the term 
$\sim \langle U \rangle +  \langle U^\dagger \rangle$ in the Lagrangian. } 
 That means that, starting from the energy
$E \sim  F_\pi $~\footnote{Well, rather from $E \sim 2\pi F_\pi \sim \mu_{\rm hadr}$.} 
(the {\it unitarity limit} for chiral theory), loop corrections to the tree scattering
amplitudes become essential. Indeed, by dimensional counting, the second and the third  terms in \p{scat-pion} 
grow with energy as $\sim E^4$. At $E \sim \mu_{\rm hadr}$, they are of the same order as the tree amplitude \p{Mtree-pion}. 
$\mu_{\rm hadr}$ is the scale where {\it all} loop corrections become equally important and perturbative calculations
are not possible anymore. 

 The author remembers well that, at the beginning of the 1970-ties, when most theorists
 did not look at the direction
of the Standard Model yet, they tried to cross the unitarity barrier in Fermi theory by 
inventing ingenious
resummation schemes \cite{Mak-Ter-Zam}. But this did not work. Actually, there is no 
much sense to calculate loops with the Fermi Lagrangian even at the energies below
the unitary limit, because it is practically impossible to separate the loop corrections 
from the contribution due to higher-derivative counterterms, 
which enter with unknown arbitrary coefficients.  
   
In effective chiral theory, loop calculations make, however,  a certain sense.
They bring about a nontrivial information 
 because of the presence of so-called {\it chiral logarithms} - large factors 
$\sim \ln (\mu_{\rm hadr}/m_\pi)$ associated with {\it infrared} divergences in massless theory, which contribute to strong amplitudes. But, again, this works only 
for low enough energies.

The conventional Einstein's  gravity is also a non-renormalizable theory. 
The Einstein-Hilbert Lagrangian reads
   \be
  \lb{Lagr-EH}
  {\cal L}^{(0)} \ =\ - \frac  1{16 \pi G_N} R \, .
   \ee 
 Newton's constant $G_N = m_{Pl}^{-2}$ carries dimension, it plays exactly the same role
as $G_F$ in Fermi theory and $F_\pi^{-2}$ in chiral theory. Tree graviton 
scattering amplitudes grow with energy. 
  
 Gravity has, however, a distinguishing feature, compared to two other
theories. In gravity, one-loop graphs are {\it free} from logarithmic divergences (power
divergences survive). Indeed, the logarithmic 1-loop divergences  should be associated with the appearance of the counterterms of canonical dimension $d=4$. 
The only such general covariant structures are 
   \be
\lb{Lgrav-d4}
{\cal L}_1^{(1)} = \ R^2, \ \ \ \ \ \ \ \ {\cal L}_2^{(1)} = \ R_{\mu\nu}
 R^{\mu\nu} 
 \ee
[the square of the Riemann tensor $R_{\mu\nu\alpha\beta}$ is expressed via a linear
combination of the structures \p{Lgrav-d4} plus a total derivative]. 
One can observe, however, that these structures {\it vanish} on the mass shell --- for the
fields satisfying the   Einstein equations 
of motion in empty space, $R_{\mu\nu} = 0$.  And hence the logarithmic 1-loop divergences cancel \cite{Hooft}. 

One can often hear people saying
 that 
``Einstein's gravity is finite at one loop''. One should clearly understand, however, that they mean thereby only the 
absence of {\it logarithmic} ultraviolet divergences 
(the absence of the analog of the second term in \p{scat-pion} for graviton scattering amplitudes). 
 But  power divergences, similar to those that show up in the first term of \p{scat-pion},
 are still there.
And the statement that, after renormalization,
 the 1-loop contribution to the amplitude is of the same order
 as the tree contribution, when the energies are of order of the Planck mass, is still there.

Going back to logarithmic ultraviolet divergences, they  reappear in gravity at the 2-loop level. 
 There exists a counterterm 
of canonical dimension $d=6$, which does not vanish on mass shell,
   \be
\lb{Lgrav-d6}
{\cal L}^{(2)} = \ R_{\mu\nu\alpha\beta} R^{\mu\nu\gamma\delta} 
R^{\alpha\beta}_{\ \ \gamma\delta} \, .
 \ee
  This brings about the contribution $\propto G_N^3 \ln \Lambda$ in the graviton-graviton
scattering amplitude.

In other words, the Einstein-Hilbert Lagrangian, like Fermi's 4-fermion Lagrangian
and like the pion interaction Lagrangian \p{Lchiral-tree}, describes well low-energy phenomena, 
but general relativity is an effective, rather than the fundamental theory.  
We will argue later that the same probably 
concerns Poincar\'e supergravity and extended Poincar\'e supergravity. No calculations
are possible in these theories beyond the Planck mass barrier.

\section{Witten's model with higher derivatives}
\setcounter{equation}0

In non-supersymmetric models, the counterterms have the same symmetry as the
tree Lagrangian --- chiral  symmetry for the effective model of pion interactions 
and general covariance for gravity. However, the situation is different in  
 supersymmetric theories --- more often than not, one cannot keep the full symmetry of the 
model in interest for individual counterterms of a given canonical dimension.
This behaviour can be  best understood 
by studying certain toy SQM models \cite{BIS}. 

The simplest possible example is Witten's SQM 
system involving one bosonic degree of freedom $x(t)$ and its fermionic superpartners $\psi(t)$ and $\bar \psi(t)$.
 The Lagrangian of the model reads 
  \be
\lb{LSQMcom}
  L_0 \ =\ \frac {\dot{x}^2 - [V'(x)]^2}2  + \frac i2 \left(
\dot{\psi} \bar \psi - \psi \dot{\bar\psi} \right)
+  V''(x) \bar \psi \psi \, .
  \ee
The corresponding equations of motion are
 \be
 \lb{eqmotSQM}
\ddot{x} + V'(x) V''(x) - V'''(x) \bar \psi \psi &=& 0 \, , \nn
i \dot{\psi} - V''(x) \psi  &=& 0 \, , \nn
i \dot{\bar \psi} + V''(x) \bar \psi  &=& 0 \, .
  \ee

The Lagrangian \p{LSQMcom}
is invariant (up to a total time derivative) under the following nonlinear supersymmetry transformations
  \be
\lb{trans_bez_D}
\delta x \equiv \delta_\epsilon x +  \delta_{\bar \epsilon} x
\ = \ \epsilon \bar \psi + \psi \bar \epsilon , \nn
\delta \psi \equiv \delta_\epsilon \psi \ =\
-\epsilon [  i\dot{x} + V'(x)] , \nn
\delta \bar\psi \equiv \delta_{\bar \epsilon} \bar\psi
\ =\ \bar\epsilon [i\dot{x} - V'(x) ] \, .
    \ee

In (0+1) dimensions, there is no need of renormalization and of introducing  higher-dimensional
counterterms. Still, one can study  relatives of the Lagrangian $L_0$, 
 having higher canonical
dimension and involving higher time derivatives, on their own merits.

The primary observation is that it is {\it impossible} to write a
Lagrangian depending on the fields $x,\psi, \bar\psi$
and involving their {\it higher} time derivatives
 which would be
invariant under the transformations \p{trans_bez_D}.
This is due to the well-known fact that the Lie
brackets of the transformations \p{trans_bez_D} do not close off mass shell,
 but only on mass shell. When acting on the variable $x(t)$, the Lie bracket
 $(\delta_{ \bar \epsilon} \delta_{\xi} - \delta_{\xi} \delta_{ \bar \epsilon})$
boils down to a total time derivative. But it is not so for the fermion variables.
For example,
   \be
\lb{Lie_bracket}
\left( \delta_{ \bar \epsilon} \delta_{\xi} - \delta_{\xi} \delta_{ \bar \epsilon}
\right) \psi =  \ \xi \bar \epsilon [i\dot{\psi} + V''(x) \psi] \ =\ 2i\xi \bar \epsilon \, \dot{\psi} +
\xi \bar \epsilon \frac {\partial L}{\partial \bar \psi} \, .
     \ee
 The presence of the  second term in \p{Lie_bracket} does not affect the
invariance of $L_0$ under
\p{trans_bez_D}. Indeed,
  \be
 \lb{Lie_L0}
 \left(\delta_{\bar \epsilon} \delta_\xi -  \delta_\xi \delta_{\bar \epsilon} \right) L_0 \ =\
  2i\xi \bar \epsilon \, \dot{L_0} + \xi \bar \epsilon \, \left( \frac {\partial L_0}{\partial \psi}
 \frac {\partial L_0}{\partial \bar\psi} +  \frac {\partial L_0}{\partial \bar \psi}
 \frac {\partial L_0}{\partial \psi} \right) = 2i\xi \bar \epsilon \, \dot{L_0} \, .
   \ee
But,  for $L \neq L_0$, the second term in the Lie bracket
$\left(\delta_{\bar \epsilon} \delta_\xi -  \delta_\xi \delta_{\bar \epsilon} \right) L$  vanishes
only on the mass shell of $L_0$.

The standard way to solve this problem and to construct fully supersymmetric actions
of any dimension
is to introduce a supervariable \p{SQM-X}. 
 The transformations of the superspace coordinates generate {\it linear} supersymmetry
transformations of the dynamic variables,
   \be
\lb{trans_with_D}
\delta x \ = \ \epsilon \bar \psi + \psi \bar \epsilon , \nn
\delta \psi \ =\ \epsilon (F - i\dot{x} ) , \nn
\delta \bar\psi \ =\ \bar\epsilon (F + i\dot{x} ) , \nn
 \delta F = i(\epsilon \dot{\bar \psi} - \dot{\psi} \bar \epsilon )\, .
    \ee

Any higher-derivative action of the form
\be
\lb{polynom}
 S \ =\ \int  dt\, d\bar\theta d\theta
\left(  \frac 12 \bar D X P\Bigl[i \frac{\partial\, }{\partial t}\Bigr] DX + V(X) \right),
  \ee
where $P(\partial/\partial t)$ is an arbitrary polynomial and
  \be
D = \frac {\partial }{\partial \theta } + i\bar \theta \frac {\partial}{\partial t}\, , \ \ \ \ \ \ \ \ \
\bar D = -\frac {\partial }{\partial \bar \theta } - i \theta \frac {\partial}{\partial t}
 \ee
are the supersymmetric covariant derivatives, is invariant under \p{trans_with_D}.

The original Witten's model \cite{Witten81} did not involve higher time derivatives, $P(z) = 1$.
In Ref. \cite{Robert}, interesting  higher-derivative models with  $P(z) = z$ and $P(z) = a + bz$
were analyzed. The component Lagrangian of the model with $P(z) = z$ reads
  \be
\lb{d1tcomp}
 L  \ =\   \dot{x} \dot{F} + \dot{\bar \psi} \dot{\psi} + V'(x) F + V''(x) \bar \psi \psi   \, .
 \ee
 By construction, it is invariant modulo a total derivative under transformations \p{trans_with_D}. 
In contrast to the Lagrangian  \p{LSQMcom}, the field $F$  now enters with derivatives. It is dynamical,
not auxiliary anymore.  

The spectrum of the corresponding 
quantum Hamiltonian  (it  now involves two pairs of dynamical variables) 
is not bounded neither from above, nor from below. The absence
of the ground state means  the presence of {\it ghosts}. Generically, ghosts bring about
   {\it collapse}: the system runs into a singularity in a finite time  (the same phenomenon 
as falling into the center for the attractive potential
$V(r) = \alpha/r^2$ with large enough $\alpha$), the probability   ``leaks through'' 
 and the unitarity is lost.

It turns out, however, that, in the particular model \p{d1tcomp}, the ghosts are ``benign'': 
there is no collapse, the Hamiltonian is Hermitian (in spite of the absence of the ground state) and 
the unitarity is preserved \cite{Robert,PU}.  

In \cite{TOE}, we conjectured that the fundamental Theory of Everything
 is not String Theory, but a conventional
quantum field theory living in a flat higher-dimensional bulk (and our Universe represents a classical
3-brane solution of this theory). For renormalizability, this theory should involve higher time and spatial
derivatives. Then it must involve ghosts, but the ghosts should be of benign variety: the 
Hamiltonian should be still Hermitian, and the $S$-matrix still unitary...
 
But in this paper, we are not interested in the dynamical properties of the higher-derivative models.
It is the fact that one cannot get rid of the former auxiliary field $F$ in this system which is 
of a principal importance for us now.

The much-studied four-dimensional supersymmetric gauge theories exhibit the same behaviour. 
 Consider first
the ${\cal N} =1, \ 4D$  supersymmetric SYM Lagrangian.
It involves the gauge fields and  gluinos and  is invariant
under certain nonlinear supersymmetry
transformations. One also  can write higher-derivative off-shell
supersymmetric Lagrangians of canonical
dimensions $d=6,8$, etc., but they necessarily include the auxiliary field $D$ of the vector
multiplet, which now becomes dynamical. In this case, supersymmetry is realized linearly.

The same is true for the  ${\cal N} =2$  supersymmetric SYM theory. We have a gauge superfield
${\cal W}$ involving a triplet of auxiliary fields $D^A$. Higher-derivatives supersymmetric
Lagrangians like ${\cal L} \sim \left\langle \int d^8\theta \, {\cal W}^2
{\bar {\cal W}}^2 \right\rangle $
can be written, and they involve  the derivatives of $D^A$. For the ``matter'' fields belonging
to the ${\cal N} =2$ hypermultiplet, the full set of the auxiliary fields is infinite. The latter can be
 presented as
components of a certain ${\cal N} =2$  harmonic
superfield \cite{HSS}.  Higher--derivative off-shell-invariant actions can also be written  in that case.

But for the ${\cal N} =4$ theory, the situation is different. Superfield formalism, with all
supersymmetries being manifest and off-shell,
is not developed, the full set of auxiliary fields is not known and probably does not exist.
Thus, one simply {\it cannot} write in this case an off-shell supersymmetric higher-derivative 
action.~\footnote{We make a terminological comment on what exactly do we mean by ``off-shell''. 
 A Lagrangian  is called off-shell symmetric 
if the corresponding action is invariant under certain variations of dynamic variables,
 without imposing any
 supplementary conditions. 
In case of supersymmetry, the action is always off-shell invariant if it can be expressed
 in terms of appropriate superfields.
But the inverse is not generally true. The ${\cal N} =4$ $4D$ SYM action 
is off-shell invariant, but cannot be expressed
into superfields.}

 In many cases, one can write, however, complicated higher-derivative  effective 
Lagrangians not involving auxiliary fields and fully  
 off-shell supersymmetric. Again, this can be best understood by considering a simple SQM  example. 
We  now consider the model with $P(z) = 1 + gz^2$ \cite{BIS}. 
In this case, the ghosts are malicious enough to bring about the collapse and to kill the Hermiticity.
But, as  we only use this as a toy model displaying the structure
of the effective Lagrangians in complicated field theories of interest, 
we need not to worry about it. 
We obtain  the following component Lagrangian,
  \be
     \lb{d2tcomp}
 L \ =\   \frac 12 (\dot{x}^2 +   F^2 ) +i \dot{\bar\psi}
    {\psi} + F V'(x) + V''(x) \bar\psi \psi  \ + g  \frac 12
\left( \ddot{x}^2 + \dot{F}^2 + 2i  \ddot{\bar\psi}
    \dot{\psi} \right).
       \ee
This Lagrangian is invariant under the transformations \p{trans_with_D}. Also in this case 
the formerly auxiliary field $F$ has become dynamical and cannot be  eliminated algebraically.
 Still, one can now integrate out the field $F$
perturbatively through the formal power series solution
   \be
F =- \sum_{n=0}^\infty g^n  \frac{d^{2n} V'(x) }{d t^{2n}}  \, .
  \ee

One obtains in this way the Lagrangian
   \be
\lb{LSQMeff}
  L  \ =\ \frac{1}{2} \bigl(  \dot{x}^2 + g \ddot{x}^2\bigr)  + i \dot{\bar\psi} \psi + i g \ddot{\bar\psi}
    \dot{\psi}
- \frac{1}{2} \sum_{n=0}^\infty (-g)^n \biggl(  \frac{d^{n}  V'(x) }{d t^{n}} 
 \biggr)^2 +  V''(x) \bar \psi \psi  \, ,
  \ee
which involves only $x, \psi$ and $\bar \psi$ and 
is by construction invariant with respect to the nonlinear supersymmetry transformations,
    \be
\lb{trans_eff}
\delta x \ &=& \ \epsilon \bar \psi + \psi \bar \epsilon \, , \nn
\delta \psi \ &=&\ \epsilon \biggl(   -i\dot{x} -  \sum_{n=0}^\infty g^n 
 \frac{d^{2n} V'(x)  }{d t^{2n}}   \biggr) \,  , \qquad
\delta \bar\psi \ =\ \bar\epsilon  \biggl( i\dot{x} -  \sum_{n=0}^\infty g^n  
\frac{d^{2n} V'(x) }{d t^{2n}}    \biggr),
    \ee
which close modulo the  equations of motion for the full Lagrangian \p{LSQMeff}. For example,
   \be 
\lb{Lie_eff}
\left(\delta_{ \epsilon_1} \delta_{\epsilon_2} -
\delta_{\epsilon_2}  \delta_{\epsilon_1}
\right) \psi = \ -2\epsilon_1 \epsilon_2  \sum_{n=0}^\infty g^n  \frac{d^{2n} \, }{d t^{2n}}  \left(
\frac {\partial L}{\partial \psi} \right).
     \ee
The Lagrangian  \eqref{LSQMeff} represents an infinite  perturbative series in $g$,
\be L = \sum_{n=0}^\infty g^n L_n = L_0 + g L_1 + g^2 L_2 + \dots \, ,
 \ee
where
$L_0$ is written in
 \eqref{LSQMcom}. 
The same is true for the supersymmetry transformations \p{trans_eff}: 
 $\delta = \delta_0 + g \delta_1 + \dots $,  with $\delta_0$ written in  \eqref{trans_bez_D}.

The full variation of the full Lagrangian, $\delta L$, should represent a total time derivative. Disregarding
the latter and expanding $\delta L \approx 0$ 
in $g$,~\footnote{$\approx$ means ``equals modulo a total derivative''.}
 we obtain in the first order in $g$,
   \be  
\lb{onshell-01}
\delta_0 L_1 + \delta_1 L_0  \approx 0
 \, . \lb{two_delta} \ee

  The variation of $L_0$ is proportional to  the classical equations of motion \eqref{eqmotSQM}.
It follows that the first-order correction to the Lagrangian,
   \be
\lb{L_1}
 L_1 =  \frac{1}{2} \ddot{x}^2  +  i \ddot{\bar\psi}
    \dot{\psi} + \frac{1}{2} \dot{x}^2 \bigl( V''(x) \bigr)^2 ,  
  \ee
is not invariant under the action of the  nonlinear
supersymmetry transformations \p{trans_bez_D} off shell, but it is invariant
(modulo  a total time derivative) {\it on shell}, when the equations of motion \eqref{eqmotSQM}
are imposed as constraints.

 On the contrary, the second-order correction
\be L_2 =  - \frac{1}{2} \bigl[ \ddot{x} V''(x) + \dot{x}^2 V'''(x)\bigr]^2, \ee
is not invariant with respect to   $\delta_0$, but satisfies a more complicated condition,
    \be
\lb{onshell-012}
\delta_0 L_2 + \delta_1 L_1 +
\delta_2 L_0  \approx 0    \, . \lb{three_delta}
 \ee

\section{Two models}
\setcounter{equation}0

The example above was simple, but somewhat artificial. However, 
the situation when the effective Lagrangian represents an infinite series of
higher-derivative terms, like in \p{LSQMeff}, and this Lagrangian is invariant under
modified supersymmetry transformations, 
 also representing an infinite series,
is quite general. 
One known example is the non-Abelian Born-Infeld effective Lagrangian \cite{Tseytlin}. It is relevant for us, 
because the leading term in the $6D$ version in this Lagrangian coincides with the Lagrangian
of ${\cal N} = (1,1)$ $6D$ SYM theory, the point of our primary interest here. 
But we consider first
  more simple  examples carrying all the salient 
features of the complicated field-theoretical and string models.

\subsection{Maximal SQM}

We consider first the so-called
{\it maximal} $N=16$ SQM obtained by dimensional reduction from 
$N=4,  \ 4D$ SYM theory or $N=1,  \ 10D$ SYM theory. It is  convenient
to describe the maximal SQM in $10D$ notations and write 
  \be
\label{L} 
L = \frac 12 \dot{A}_I^A \dot{A}_I^A 
- \frac {g^2}4 
f^{ABE} f^{CDE} A^A_I A^B_J A^C_I A^D_J + \frac i2 \lambda_\alpha^A 
\dot{\lambda}_\alpha^A  -
\frac{ig}2 f^{ABC} \lambda^A_\alpha (\Gamma_I)_{\alpha\beta} \lambda_\beta^B
 A^C_I \ ,
\ee
$I= 1,\ldots, 9$, $\lambda^A_\alpha$ are real fermions, 
$\alpha = 1, \ldots, 16$,  and $ (\Gamma_I)_{\alpha\beta}  =  (\Gamma_I)_{\beta\alpha}$ 
are the $10D$ gamma matrices, $\Gamma_I \Gamma_J + \Gamma_J \Gamma_I = 2\delta_{IJ}$. 

We consider the simplest such model with $SU(2)$ gauge group. 
As is well known, the non-Abelian field strength $\hat F_{IJ}  = [\hat A_I, \hat A_J]$ 
and the quartic classical potential $V \sim \langle (\hat F_{IJ} )^2 \rangle $  vanish in the  
{\it Abelian valley},
  \be
  A^A_I = A_I c^A \, ,   
  \ee
 It is also known that the bottom of 
this valley is not lifted by quantum corrections so that 
the system tends to spread along the valley. In the region $g|\vecg{A}|^3  \gg 1$, 
one can evaluate the effective Lagrangian depending only on the slow variables
$A_I$ and its superpartners in the Born--Oppenheimer  framework as a series
over the  small BO parameter, 
  \be
\lb{BO}
 \gamma \ =\ \frac 1{g|\vecg{A}|^3} \, . 
  \ee
  To the lowest order in $\gamma$, the Lagrangian is just
  \be
\lb{Ltreemax}
L_0 \ =\ \frac 12 \dot{A}_I^2 + \frac i2 \lambda_\alpha \dot{\lambda}_\alpha \, .
  \ee
It is invariant under supersymmetry transformations
  \be
\lb{trans_N=16}
  \delta_0 A_I = -i \epsilon \Gamma_I \lambda , \ \ \ \ \ \ \ \ 
\delta_0 \lambda = \dot{A}_I \Gamma_I \epsilon \, ,
  \ee
It is not so difficult to prove that the action based on the 
free Lagrangian \p{Ltreemax} is the {\it only} one  invariant under \p{trans_N=16} \cite{PSS}. 
The effective BO Lagrangian involves, however, many other nontrivial terms. 
 As was mentioned,  no potential is generated. Also, there are no corrections 
to the metric (quadratic in derivative terms). The first relevant correction 
is quartic in derivatives,
  \be
\lb{Lv4r7}
 L_1 \ \propto \ \frac {(\vecg{E}^2)^2}{g^3|\vecg{A}|^7 }\,  + {\rm fermion \ terms} \, ,
 \ee 
($\vecg{E} = \dot{\vecg{A}}$). 
As was promised in the Introduction, it is suppressed as $\gamma^3$, compared to \p{Ltreemax}.~\footnote{It 
is best seen in the Hamiltonian language, where the characteristic values of the canonical  momentum 
$\vecg{E}$ for the low-lying levels are $E_{\rm char} \sim A_{\rm char}^{-1}$.}

\begin{figure} [t]
\begin{center}
\includegraphics[width=2in]{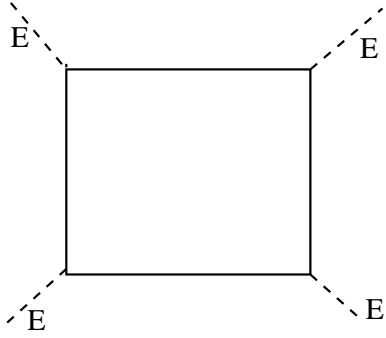}
\end{center}
\caption{The graph giving the effective Lagrangian \p{L4box}. 
Solid lines are the propagators of the ``fast'' charged modes, $A_I^\pm$. }
\label{Box}
\end{figure}

 The bosonic part \p{Lv4r7} of $L_1$ can be determined by  evaluating of the graph
in Fig.~\ref{Box}. 
 Using the background field formalism, one derives
  \be
\lb{L4box}
 L_1  \propto g^4(\vecg{E}^2)^2 \int_{-\infty}^\infty  \frac {d\omega}{(\omega^2 + g^2 \vecg{A}^2)^4} \, ,
  \ee
which gives \p{Lv4r7}.
But to restore all other terms involving $\lambda_\alpha$ is a rather nontrivial task. 
The full expression was derived in \cite{japoncy}.
We will not reproduce it here.

The contribution $S_1$ in the effective BO action is not invariant under \p{trans_N=16}.
On the other hand, the whole series \p{L-series} represents the effective BO Lagrangian for the original
non-Abelian model \p{L}. The latter has the full off-shell ${\cal N} = 16$ supersymmetry. 
The effective Lagrangian also must have it. And it has, only the supersymmetry transformations 
are now {\it modified} by the same token as they were in the simplest SQM example considered in the previous
section. The modified transformations represent an infinite series  \p{delta-series} 
with certain $\delta_1, \delta_2$ etc. A (rather complicated) expression for $\delta_1$ can be found 
in \cite{japoncy}. 

Expanding $\delta L \approx 0$ in $\gamma$ and keeping the terms $\sim \gamma^3$, we derive the  relation
\p{onshell-01}.
  This means that, even though the Lagrangian $L_1$ is not invariant under $\delta_0$ 
 off shell, it {\it is} invariant on shell ---
 with taking the equations of motion of the Lagrangian \p{Ltreemax} into account.

It is instructive to check it explicitly.
The equations of motion have in this case a very simple form,
 \be
\lb{eqmotLmax}
  \ddot{\vecg{A}} \equiv \dot{\vecg{E}} \ =\ 0, \ \ \ \ \ \ \dot{\lambda_\alpha} = 0 \, .
 \ee
If one neglects in $L_1$ the terms that vanish on mass shell
 [vanish under the conditions \p{eqmotLmax}], 
the expression for $L_1$ greatly simplifies. In \cite{Plefka}, 
it was expressed as 
  \be
\lb{Plefka}
 L_1^{\rm mass \ shell}  \ \propto \
 \frac {(\vecg{E}^2)^2}{|A_I  - i (E_J/2\vecg{E}^2) \, \lambda \Gamma^{IJ} \lambda|^7 } \, ,
   \ee
where  $\Gamma^{IJ} = \Gamma^{[I} \Gamma^{J]}$. 
 The variation of \p{Plefka} under \p{trans_N=16} amounts to a total time derivative
 plus the terms involving
$\ddot{\vecg{A}}$ and $\dot{\lambda}_\alpha$. Indeed,  it is easy to see that, 
under the conditions \p{eqmotLmax},
    \be
\lb{var-znam}
\delta_0 \left( A_I  - i \frac {E_J}{2\vecg{E}^2}  \, 
\lambda \Gamma^{IJ} \lambda \right) \ =\ 
\frac {iE_I E_K}{\vecg{E}^2} \, \lambda \Gamma^K \epsilon \, .
   \ee
Then
  $$
\delta_0  L_1^{\rm mass \ shell}  \ =\ f(\vecg{E}, \lambda_\alpha) E^I \frac 
\partial {\partial A_I} g(\vecg{A}, \vecg{E}, \lambda) 
 $$
with some $ f, g$. Using again \p{eqmotLmax}, this gives
  \be
\delta_0  L_1^{\rm mass \ shell}  \ \ \stackrel{on\ mass\ shell}{\equiv}  \ \ f \frac {d}{dt} g \ \
\stackrel{on\ mass\ shell}{\equiv} \ \ \frac {d}{dt} (fg) \, .
  \ee

The bosonic effective Lagrangian involves also the terms with still higher derivatives. 
For example, there is a
term
  \be
\lb{L6maxbos}
   L_2 \propto \ \frac {(\vecg{E}^2)^3}{|\vecg{A}|^{14}}
  \ee
that is suppressed as $\gamma^6$, compared to \p{Ltreemax}. The coefficient in \p{L6maxbos} 
was evaluated in   an accurate 2-loop calculation in \cite{Becker}. The full 
expression for $L_2$ including fermions is not known.
 Neither is known the simplified expression for $L_2$ 
disregarding the terms proportional
to \p{eqmotLmax}.

  Note that this expression {\it does not} need to be invariant with respect to 
\p{trans_N=16} on mass shell and 
is probably  not. Indeed, keeping the terms of order $\sim \gamma^6$, we can only derive
the relation \p{onshell-012}.
   The last term there 
is proportional to the equations of motion \p{eqmotLmax}. But the term $ \delta_1 L_1 $ is not, 
and one cannot make any conclusions about $\delta_0 L_2 $. Probably, the true $L_2$, 
entering the true BO effective
Lagrangian, is not on-shell supersymmetric,
even though a supersymmetric on-shell invariant  can in principle be written:
   \be
\lb{quasi_Plefka}
 {\tilde L}_2  \ \propto \ \frac {(\vecg{E}^2)^3}{|A_I  - i (E_J/2\vecg{E}^2)  \,
\lambda \Gamma^{IJ} \lambda|^{14} } \, .
   \ee

\subsection{Tree chiral Lagrangian}

We now go back to the effective chiral theory, being not concerned this time with loop corrections,
but only with the structure of the leading term \p{Lchiral-tree} and of the tree amplitudes that it
 generates.\footnote{I am indebted to G.~Bossard who attracted my attention
to this example.}

Expanding the exponentials, we can present the Lagrangian \p{Lchiral-tree} as an infinite series, 
$$ {\cal L}^{(0)} \ =\ {\cal L}_0 + {\cal L}_1 + \ldots \, ,$$
the beginning of this series being written in \p{Lchiral-expan}. The full Lagrangian is invariant under
the transformations,
  \be
\lb{Utran-LR}
U \ \to \ \exp\{i\sigma^a \alpha^a \} U, \ \ \ \ \ \ \ \ \ \ \ \ \ 
U \ \to \ U \exp\{i\sigma^a \beta^a \}\, .
   \ee
Consider e.g. the left multiplication. Infinitesimally, it gives
  \be
\lb{del-al-phi}
\delta_\alpha \phi^a \ =\ - \varepsilon^{abc} \alpha^b \phi^c + 
\frac {|\vecg{\phi}|}{{\rm tg} |\vecg{\phi}|} \alpha^a  + \frac {\vecg{\alpha} \cdot 
\vecg{\phi}}{|\vecg{\phi}|^2} \left[ 1 -   \frac {|\vecg{\phi}|}{{\rm tg} |\vecg{\phi}|} \right] \phi^a
 \,  ,
  \ee
where $\phi^a = \pi^a/F_\pi$. 
If also $\vecg{\phi}$ is small, this can be presented as 
 \be
\lb{del-al-phi-ser}
\delta_\alpha \phi^a =    - \varepsilon^{abc} \alpha^b \phi^c + \alpha^a +  
\frac 13 \alpha^b (\phi^b \phi^a -
\delta^{ab} \vecg{\phi}^2 ) + O(\phi^4)  \ \equiv \   - \varepsilon^{abc} \alpha^b \phi^c + 
\delta_0 \phi + \delta_1 \phi + \ldots\, .
  \ee
The first term in \p{del-al-phi-ser} describes the $SO(3)$ rotations, the diagonal symmetry 
$U \to \Omega U \Omega^\dagger$, with respect to which all terms of the expansion \p{Lchiral-expan} are
still invariant. But it is not true for other contributions in \p{del-al-phi-ser}.
Actually, we may now observe that the leading term in the expansion is still invariant under the 
translations $\delta_0\phi^a  = \alpha^a$. In addition, we may observe that the variation 
 $\delta_0 {\cal L}_1$ amounts to a total derivative, if taking into account the tree equations of
motion $\Box \phi^a = 0$. In other words, $\delta_0 {\cal L}_1$ vanishes on mass shell!
 In the full analogy with the maximal SQM model and with the SQM model of Sect.~3, this is a direct
corollary of the fact that the full series in \p{Lchiral-expan} 
is invariant under the full series in \p{del-al-phi-ser}.

Note that the term $\sim \pi^6$ in the expansion (we need not to write it explicitly) 
is {\it not} invariant on shell under  translations, it only satisfies the  condition
\p{onshell-012}. This means in particular that the tree 6-point amplitude generated by ${\cal L}_2$
does not exhibit the full chiral symmetry $SU_L(2) \times SU_R(2)$, it only keeps its diagonal 
$SO(3)$ part. On the other hand, the {\it full} tree amplitude, two relevant contributions to which
being depicted in Fig.~3, is of course   $SU_L(2) \times SU_R(2)$ - symmetric. 
\begin{figure} [t]
\begin{center}
\includegraphics[width=4in]{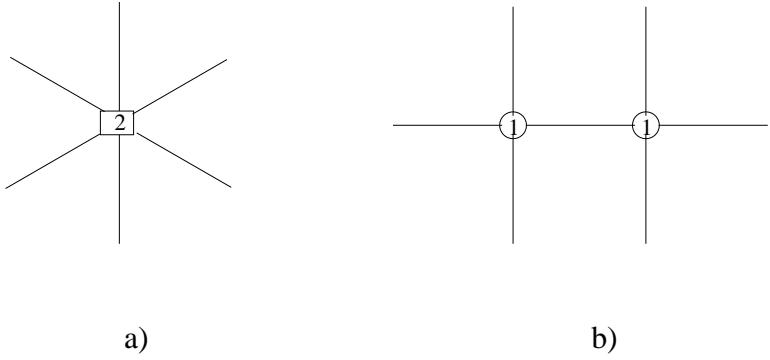}
\end{center}
\caption{Contributions of ${\cal L}_1$ and 
 ${\cal L}_2$ to 6-point amplitudes. }
\label{6-hvostka}
\end{figure}

\section{Six-dimensional gauge theories}
\setcounter{equation}0
 
In $5+1$ Minkowski space, left--handed spinors $\lambda^a$ 
and right--handed spinors $\psi_a$ ($a = 1,2,3,4$) 
belong to the different completely independent  spinor representations  (1,0) and (0,1) 
of $Spin(5,1)$.~\footnote{In Euclidean space, two spinor representations
of $Spin(6) = SU(4)$ are complex conjugate to one another. 
The situation is exactly opposite to that in four dimensions, where two
spinor representations are conjugate to one another in Minkowski space, but are
completely independent in Euclidean space.}   
  We introduce six $4 \times 4$ matrices  $\gamma_M$  
(the $6D$ analogs of $\sigma_\mu$), which  are  real, antisymmetric  
  and  satisfy
 \be
\gamma_M \tilde \gamma_N + \gamma_N \tilde \gamma_M = \ -2\eta_{MN}\,, \qquad  \eta_{MN}={\rm diag} (1,-1,-1,-1,-1,-1) \,,
 \ee
with
 \be
(\tilde \gamma_M)^{ab} \ =\ \frac 12 \varepsilon^{abcd}   (\gamma_M)_{cd} \, .
  \ee

    The minimal ${\cal N} = (1,0)$ supersymmetric 
Yang-Mills theory in 6 dimensions 
includes the gauge potential $A_M$ and a couple of 
left-handed fermion fields $\lambda^a_j$ ($ j = 1,2$) 
satisfying the pseudoreality condition,
\be
\overline{\lambda^a_j}\equiv -C_{\ b}^a(\lambda^b_j)^* = \lambda^{aj} \equiv 
\varepsilon^{jk} \lambda^a_k\, ,
\ee
where $C$ is the charge conjugation matrix with the properties
 $C = -C^T, \ C^2 = -1$. In addition, the gauge multiplet involves 
 the triplet of auxiliary fields ${\cal D}_{jk} = {\cal D}_{kj}$. All the fields 
represent Hermitian colour
matrices. 
 
 Being expressed in components, the 
 Lagrangian reads
 \be
\lb{gauge-6D}
 {\cal L}^{\rm gauge} \ = \ \frac 1{2f^2} \left\langle - 
F_{MN}^2 
+ i \lambda^k \gamma^M \nabla_M \lambda_k  -  {\cal D}_{jk} {\cal D}^{jk} 
\right\rangle \, ,   
  \ee
with $\nabla_M X = \partial_M X - i [A_M,X],  \ F_{MN} = [\nabla_M, \nabla_N]$. 
The constant $f^2$ carries the dimension $m^{-2}$, the same as Newton's 
gravity constant and as $F_\pi^{-2}$.

As this theory includes only left-handed fields, it involves a chiral anomaly 
\cite{anom},
which breaks gauge invariance.
 To compensate this anomaly, one should add an adjoint matter hypermultiplet, 
involving the right-handed pseudoreal fermions $\psi^{A=1,2}_a$ 
and four real scalars $\varphi_\alpha$. The corresponding Lagrangian reads
  \be
\lb{hyper-6D}
  {\cal L}^{\rm hyper} \ = \ \frac 1{2f^2} \left\langle    
- i \psi^A {\tilde \gamma}^M \nabla_M \psi_A   
+  (\nabla_M \varphi_\alpha)^2  -   \psi^A_a [\lambda^{ka}, \varphi_{kA}] - 
 \varphi^{kA} \varphi^l_{A}\,
 {\cal D}_{kl}  \right\rangle
  \ee
 with~\footnote{Note the constraint $\varphi^{kA} = -(\varphi_{kA})^*$  that the fields \p{phi-kA}  
satisfy.} 
 \be
\lb{phi-kA}
 \varphi^{kA}  = \frac 1{\sqrt{2}} (\sigma_\alpha)^{kA} \varphi_\alpha \ =\ \frac 1{\sqrt{2}} \left( 
\begin{array}{cc} i\varphi_0 + \varphi_3 & \varphi_1 - i\varphi_2
\\ \varphi_1 + i\varphi_2 & i\varphi_0 - \varphi_3 \end{array} \right) \, . 
  \ee  
To provide for the extended supersymmetry of the
sum $ \cL^{(1,1)} = {\cal L}^{\rm gauge} + {\cal L}^{\rm hyper}$, 
the constant $f$ in \p{hyper-6D} is chosen to be the same as in \p{gauge-6D}.
After excluding the auxiliary fields 
${\cal D}_{jk}$,  we obtain
  \be
\lb{L-11}
 \cL^{(1,1)}  = \frac 1{2f^2} \left\langle    - 
F_{MN}^2 
+ i \lambda^k \gamma^M \nabla_M \lambda_k 
- i \psi^A {\tilde \gamma}^M \nabla_M \psi_A   
+  (\nabla_M \varphi_\alpha)^2  \right. \nn
\left. - \psi^A_a [\lambda^{ka}, \varphi_{kA}]  + \frac 1{16} 
[ \varphi^{kA} , \varphi^l_{ A}] [  \varphi^{B}_k , \varphi_{lB} ]
 \right\rangle
  \ee
One can be convinced that the corresponding action is invariant, indeed, under 
certain nonlinear   
 $\cN = (1,1)$ supersymmetry transformations 
with left-handed and right-handed Grassmann
parameters.

In contrast to Witten's models of Sect.~2, to $\cN = 1$ and $\cN = 2$ $4D$ 
SYM theories, but
similar to the maximal SQM model of Sect.~3 and to $\cN = 4$ $4D$ SYM 
theory, a linear 
superfield
realization of this supersymmetry is not known and probably does not exist.

As a result, one cannot write an off-shell 
$\cN = (1,1)$ invariant of
a fixed  canonical dimension $d > 4$. 
On the other hand, higher-dimensional structures that possess
$\cN = (1,1)$ supersymmetry {\it on shell} [i.e. 
the structures  invariant under
 the  $\cN = (1,1)$ supertransformations 
when the fields are constrained to satisfy the equations
of motion of the basic Lagrangian \p{L-11}] do exist. 
Such a restricted on-shell invariance
is {\it good enough} for these structures to play the role
of counterterms for the theory \p{L-11} and to give rise to logarithmic
divergences in its on-mass-shell scattering amplitudes.~\footnote{One can also 
mention here  
the existence of nontrivial higher-derivative  actions enjoying on-shell (but not off-shell) 
${\cal N} =4$
supersymmetry in four dimensions (see, e.g., \cite{Buchbinder,Drummond}). But 
these invariants are not relevant in perturbative calculations, they do not 
appear as counterterms
for the renormalizable, and even finite $\cN = 4$ $4D$ theory.}

 \subsection{Harmonic superspace and harmonic superfields}

There is no off-shell $\cN = (1,1)$ superfield formalism, but the $\cN = (1,0)$ off-shell 
superfields
well exist, can be studied and can be used. Explicit expressions for superfields realizing
irreducible representations of the supersymmetry algebra can be best written in the framework
of the harmonic superspace approach \cite{HSS}. We address the reader to our recent paper \cite{BIS} 
for a detailed description of this formalism, as applied to 6-dimensional theories, and quote
here only its salient features.

The standard ${\cal N}{=}(1,0)$ superspace involves the coordinates
\be
z=(x^M, \theta^{ai}),  \label{Centr}
\ee
where $\theta^{ai}$ are Grassmann pseudoreal left-handed spinors.

Next we introduce the harmonics $u^{\pm i}$  [$u^-_i = (u^+_i)^* , u^{+i} u_i^- = 1$], 
which describe the ``harmonic sphere'' $SU(2)_R/U(1)$,
where $SU(2)_R$ is R-symmetry group of the ${\cal N}=(1,0)$ Poincar\'e superalgebra. We now 
consider the projections $\theta^{\pm a}=u^\pm_k\theta^{ak}$ and
 introduce the ``analytic coordinate''
 $ x^M_{({\rm an})}=x^M+ \frac i2 \theta^+\gamma^M \theta^- $.

A very important property is that the set of coordinates
\be
\zeta :=(x^M_{({\rm an})}, \theta^{+a}, u^{\pm i}) \,, \label{AnalSS}
\ee
involving only a half of the original Grassmann coordinates
forms a subspace closed under the action of ${\cal N}{=}(1,0)\ 6D$ supersymmetry.
The set \p{AnalSS} parametrizes what is called ``harmonic analytic superspace''.

Many relevant superfields are {\it Grassmann-analytic} or {\it G-analytic},
 which means that they do not 
depend in the analytic basis \p{AnalSS} on $\theta^-$, but only on $\theta^+$. This makes a
series over $\theta$ much shorter and much more handleable than for a generic superfield.
G-analytic superfields are quite analogous to habitual chiral superfields in ordinary 
$4D$ $\cN = 1$ superspace, which depend either only on $\theta$ or only
 on $\bar \theta$.

It is convenient to define the  differential operators $D^+$ and 
$D^{\pm\pm}$ called
spinor and harmonic derivatives. In the analytic basis, they are expressed as
    \be
 D^+_a \ =\ \frac \partial {\partial \theta^{-a}}\,, \ \ \ \ \ 
 D^{\pm\pm}=  u^{\pm i} \frac {\partial }{\partial u^{\mp i}} + 
\frac i2 \theta^{\pm a} 
\gamma^M \theta^{\pm b} \frac {\partial}{\partial x^M} + 
\theta^{\pm a}  \frac \partial {\partial \theta^{\mp a}} \, .
 \ee

We also define the operator of harmonic charge,
 \be
 D^0 = u^{+i} \frac {\partial}{ \partial u^{+i}} -
u^{-i} \frac {\partial}{ \partial u^{-i}} + \theta^{+a}  
\frac \partial {\partial \theta^{-a}} -  \theta^{-a} 
 \frac \partial {\partial \theta^{+a}}
 \ee
and classify the superfields by the eigenvalues of $D^0$. The superfield with eigenvalue 1 will be
denoted as $X^+$, the superfield with eigenvalue $-2$ as $Y^{--}$, etc.

A supersymmetric action can be obtained by integrating a generic superfield of zero harmonic
charge over the whole superspace 
 or by integrating an analytic superfield of harmonic charge 
+4
over the analytic superspace. The corresponding measures
will be denoted 
 \be
\lb{measures} 
dZ =  d^6x \, d^8 \theta \, du \, , \ \ \ \ \ \ \   d\zeta^{(-4)} = d^6x_{\rm an} \, d^4 \theta^+ \, du\,, 
  \ee
where $du$ is the measure on the harmonic sphere (with the normalization $\int du =1$), 
$d^8 \theta = d^4 \theta^+ d^4 \theta^-$ and we have chosen the convention 
$\int d^4 \theta^+ (\theta^{+a} \theta^{+b} \theta^{+c} \theta^{+d}) \ =\ -\varepsilon^{abcd}$.

The gauge supermultiplet is described by a G-analytic 
superfield $V^{++}$.
 Its component expansion
in the Wess-Zumino
gauge is very simple, 
  \be
 \lb{V++WZ}
V^{++} \ =\ \frac 12 \theta^{+} \gamma^M \theta^{+} A_M - \frac 13 
\varepsilon_{abcd}  \theta^{+b}  \theta^{+c} \theta^{+d}  \lambda^{-a}  
+ \frac 18 \varepsilon_{abcd}  \theta^{+a}  \theta^{+b}  \theta^{+c}  \theta^{+d}
 {\cal D}^{--} \, ,
 \ee 
where ${\cal D}^{--} = {\cal D}^{jk} u_j^- u^-_k $.

Given the superfield $V^{++}$, one can also define the
 superfield  $V^{--}$ from the requirement that the commutator
 $[\nabla^{++}, \nabla^{--}]$ of the 
covariant harmonic 
derivatives, 
\be
 \lb{cov-harm}
\nabla^{++}  = D^{++} + V^{++},\ \ \ \ \ \nabla^{--} =  D^{--} + V^{--}\, ,
  \ee   
is the same as $[D^{++}, D^{--}] = D^0$.
In contrast to the G-analytic $V^{++}$, $V^{--}$ is a generic superfield.
In the following, we will also need the superfields
 \be
\lb{W+F}
 W^{+a}=-\frac{1}{6}\varepsilon^{abcd}D^+_b D^+_c D^+_d \, V^{--}\, , \ \ \ \ \ 
F^{++} =  \frac 14 D^+_a W^{+a} = 
-\frac{1}{24}\varepsilon^{abcd} D^+_aD^+_b D^+_c D^+_d \, V^{--}\, .
 \ee
 The superfield $F^{++}$ is G-analytic.

The minimal $\cN = (1,0)$ SYM action is described via $V^{++}$ as follows
  \cite{Zupnik},
   \be
\lb{act-Zup}
S^{SYM} =\frac{1}{f^2}\sum\limits^{\infty}_{n=2} 
\frac{(-1)^{n}}{n} \left\langle  \int
d^6\!x\, d^8\theta\, du_1\ldots du_n \frac{V^{++}(z,u_1 )
\ldots V^{++}(z,u_n ) }{(u^+_1 u^+_2)\ldots (u^+_n u^+_1 )}  \right\rangle 
\,,
   \ee
Substituting there \p{V++WZ}, 
we can reproduce \p{gauge-6D}. Note that the superfield
equations of motion for the action \p{act-Zup} are extremely simple,
  \be
\lb{F++0}
 F^{++} \ =\ 0 \, .
 \ee 

To describe the supermultiplet interactions, we have to introduce the 
preudoreal 
G-analytic superfield $q^{+A}$.
 The minimal $d=4$ Lagrangian \p{hyper-6D} follows
from the superfield action
   \be
\lb{act-hyp-superfield}
  S \ =\ - \frac 1{2f^2} \left \langle
 \int d\zeta^{-4} q^{+A} \nabla^{++} q^+_A  \right \rangle\, .
  \ee  
In the free case, $\nabla^{++} \to D^{++}$ and the field $q^{+A}$ satisfies 
the equations
of motion $D^{++} q^{+A} = 0 $. They can be resolved to obtain
  \be
\lb{q+free}
  q^{+A} \ =\ \varphi^{+A} - \theta^{+a} \psi_a^A - \frac i2 
\theta^{+a} \gamma^M \theta^{+b} \partial_M \varphi^{-A} \, ,
 \ee
 where  $\varphi^{\pm A} = \varphi^{kA} u^\pm_k$, with $ \varphi^{kA}$ being
 the  physical harmonic-independent pseudoreal
 [see the footnote to Eq.\p{phi-kA}]
{\it on-shell} scalar fields 
 satisfying the d'Alembert equation $\Box \varphi = 0$, and 
$\psi_a^A$ are the physical
right-handed  on-shell pseudoreal
 fermionic fields satisfying the free Dirac equation.

Generically, $q^{+A}$ involves an infinite number of other component fields.
 It turns
out, however, that, when one is only interested in the minimal $d=4$
hypermultiplet action, they  all enter the Lagrangian without derivatives and
can be algebraically excluded. 

The Lagrangian \p{hyper-6D} is obtained by substituting \p{q+free}, 
\p{cov-harm} and
\p{V++WZ} in \p{act-hyp-superfield}. 

One can be convinced that the sum of the gauge action \p{act-Zup} 
and the hypermultiplet action \p{act-hyp-superfield}
is invariant under the following $\cN = (0,1)$ supersymmetry transformations with right-handed pseudoreal
Grassmann parameters $\epsilon_{aA}$,
  \be
\lb{trans-01}
  \delta_0 V^{++} &=& \epsilon^{+A} q^+_A \, , \nn
  \delta_0 q^+_A &=& - (D^+)^4 (\epsilon^-_A V^{--} ) 
  \ee
($\epsilon^\pm_A = \epsilon_{aA} \theta^{\pm a}$). The $\cN = (1,0)$ symmetry is of course
 manifest.

\section{Effective Lagrangian and counterterms}
\setcounter{equation}0

As was mentioned in Sect.~2, in order to perform practical calculations in effective chiral theory, one should
include in the bare Lagrangian an infinite number of counterterms with UV-divergent coefficients. These
divergences cancel order by order, while calculating the loops. However, the $6D$ theory described in the 
previous section does not have practical experimental applications. We do not really want to calculate the
amplitudes, but only wish to elucidate the structure of UV divergences in this theory. It is
 then convenient for us
to assume that the bare Lagrangian is just \p{L-11}. The effective Wilsonean Lagrangian has then an explicit
dependence on the ultraviolet cutoff $\Lambda$. We will be interested 
in the higher-derivative contributions
in this effective Lagrangian. Having not found another good word, we will
 still call them "counterterms".

It is more or less clear that, when presented in such a way,  the full Wilsonean 
effective Lagrangian for this theory 
should possess the 
same symmetry, the full
${\cal N} = (1,1)$ supersymmetry, as the tree Lagrangian. Let us justify this important
claim at the {\it physical} level of rigour.
  \begin{itemize} 
\item We note first that the Wilsonean effective Lagrangian has the same nature as any other
effective Lagrangian --- it is obtained by integrating out high-momenta and high-energy modes.
 \item Consider the corresponding effective Hamiltonian. Its spectrum should match the low-energy
spectrum of the original Hamiltonian and enjoy, in particular, the same degeneracies. 
 \item But if the symmetry is preserved at the Hamiltonian level, this should also be the case for the
Lagrangian.
  \end{itemize}

In the SQM examples considered before, we have seen, however, that one cannot keep the full
symmetry for the contributions to the effective Lagrangian of a given canonical dimension, if
there is no linear (superfield) realization of the full supersymmetry or if this realization is not
implemented. Thus, one cannot expect the higher-dimensional counterterms that are
 relevant for calculations of on-shell scattering
amplitudes in the theory \p{L-11}  to have the full $\cN = (1,1)$ supersymmetry. 

However, the counterterms should be gauge-invariant. In addition, they should  enjoy 
the off-shell 
${\cal N} = (1,0)$ supersymmetry. These constraints follow from the fact that 
 one-particle-irreducible 
amplitudes calculated at a given loop order should satisfy the Ward identities following from
the gauge invariance and $\cN = (1,0)$ supersymmetry. 

The first statement (about gauge invariance) is rather common and does not require special 
comments.~\footnote{It is important, of course, that the theory \p{L-11} is anomaly-free.} The requirement
that $\cN = (1,0)$ supersymmetry is preserved follows from the existence of  $\cN = (1,0)$  superspace and superfield
description, where supersymmetry is realized linearly. In such cases, one can develop a supergraph technique
 that keeps the  $\cN = (1,0)$  supersymmetry by construction. Such a technique has not been explicitly formulated
yet,
but it should be similar in spirit to the $\cN = 2$  $4D$ 
supergraph technique described in \cite{HSS}.

Anyway, we assume that explicit perturbative calculations in \cite{Bern-gauge} manifestly 
keep this symmetry at each loop order, even though they are not supergraph calculations. This claim
is confirmed by the {\it results} obtained there.

We are now going to show 
 that the relevant counterterms do not arise at the 1-loop and 2-loop level.

\subsection{$d=6$}

Consider first possible 1-loop counterterms. Their canonical dimension should be $d=6$. 
The only gauge-invariant $\cN = (1,0)$ supersymmetric $d=6$ action involving the gauge supermultiplet $V^{++}$ 
is \cite{ISZ}
 \be
\lb{F++2}
  S^{\rm gauge}_{d=6} \ \sim \ \left \langle \int d \zeta^{-4} (F^{++})^2 \right\rangle \, ,
  \ee
with the G-analytic superfield $F^{++}$ defined in \p{W+F}. However, the equations of motion for the pure gauge theory
\p{act-Zup} are exactly $F^{++} = 0$, i.e. the action \p{F++2} vanishes on mass shell and is irrelevant. 
  
If we include the hypermultiplet, the equations of motion are modified to 
  \be
\lb{eqmot-hyper}
F^{++} + \frac 12 [q^{+A}, q^+_A] \ =\ 0 \, .
  \ee 
The pure gauge action \p{F++2} does not vanish on mass shell anymore, A generic $d=6$ action 
represents \cite{IShyper} a linear combination of \p{F++2}, of the structure 
 \be
\lb{quart}
 S_{\rm quart} \sim \ \left\langle \int d\zeta^{-4} \, [q^{+A}, q^+_A]^2 \right \rangle
 \ee
and of an infinite series of structures
  \be
\lb{Sn}
 S_n \ \sim \ \left \langle \int dZ \, q^{+A} (\nabla^{--})^n (\nabla^{++})^{n-1} q^+_A  \right\rangle \, . 
   \ee

 A generic linear combination does not vanish on the mass shell \p{eqmot-hyper}. 
We have seen, however, that the full extended supersymmetry of the full effective Lagrangian 
$L_0 + L_1 + \ldots$ implies that the tree-level supersymmetry variation  $\delta_0 L_1$ 
of the next-to-leading term 
vanishes on mass shell [see the discussion after Eq.\p{onshell-01}]. 
In our case, $L_0$ is the tree action,  $\delta_0$ are the transformations
\p{trans-01}
and we are studying the question if $L_1$ might have canonical dimension $d=6$.  
 One can show \cite{BIS} that the requirement for $\delta_0 L^{d=6}$
 to vanish on mass shell leads to the
conclusion that $L^{d=6}$ vanishes on mass shell  itself. 
Which means that
 logarithmic divergences are absent at the 1-loop
level.

\subsection{$d=8$}

We go over to two loops. Consider first possible $\cN = (1,0)$ off-shell supersymmetric counterterms of canonical dimension
$d=8$ in the pure gauge sector. One can show that all of them vanish on mass shell.~\footnote{This has been known since
\cite{HoweStelle}, but we address the reader to \cite{BIS} for a simple ``harmonic'' proof.} 

The analysis of the structures
including the hypermultiplet was performed in Ref.\cite{BIS}. At the first step, one can show that all the possible structures
can be reduced on shell to 
  \be
\lb{quart-d8}
 S^{d=8}_{\rm quart} \ \sim \ \left\langle \int dZ \, [q^{-A}, q^-_A] [q^{+B}, q^+_B] \right\rangle \, ,
  \ee
where $q^{-A} = \nabla^{--} q^{+A}$. 

At the second step, one can calculate the variation of \p{quart-d8} under the transformations \p{trans-01} and find
out that this variation does not vanish on shell. Hence, the corresponding Lagrangian does not satisfy 
the requirement \p{onshell-01} and cannot represent a next-to-leading term ${\cal L}_1$ in the Wilsonean effective Lagrangian. 
Note that we used in this reasoning the already established fact that there are no $d=6$ counterterms and the first  
correction (if any) should start from $d=8$. 

On the other hand, if we lift the requirement of off-shell $\cN = (1,0)$ invariance, nontrivial $d=8$ invariants
possessing  on-shell $\cN = (1,0)$ and $\cN = (1,1)$ symmetries
 can be written. Again, consider first the
pure gauge sector. We can write \cite{BP,BIS} 
  \be
\lb{W4-single}
 S^{d=8}_1 \ \sim \ \left\langle \int d\zeta^{-4} \, \varepsilon_{abcd} \, W^{+a} W^{+b} W^{+c} W^{+d} \right\rangle \, .
  \ee
This action is not $\cN = (1,0)$ invariant, because $W^{+a}$ is not a G-analytic superfield, $D^+_a W^{+a} = 4F^{++} \neq 0$, 
and the
integrand in \p{W4-single} is not a G-analytic superfield. But \p{W4-single} is  $\cN = (1,0)$ invariant {\it on shell}. 
Indeed, in the pure gauge case, $F^{++}$ vanishes on shell and hence $W^{+a}$ is G-analytic on shell!

Being expressed through components, the bosonic part of \p{W4-single}
gives the known $ F^4$ structure \cite{Gross} (see \cite{Bergshoeff} for the complete component form),
  \be
\label{F4}
 [{\cal L}^{d=8}_1]_{\rm bos} &\sim & \left \langle 
  2
F_{MN} F^{MN} F_{PQ} F^{PQ}  +
F_{MN} F_{PQ} F^{MN} F^{PQ} \right. \nn
&&
\left.  {\;\;\;} -\, 4F^{NM} F_{MR} F^{RS} F_{SN}  - 8
F^{NM} F_{MQ} F_{NR} F^{RQ}  \right \rangle_{(s)}.
 \ee
 Note the presence of the  symmetrized
color trace $\langle  \rangle_{(s)} \sim \left\langle  X(Y
Z U  + UYZ + ZUY) \right\rangle $ in \p{F4}.
 
The invariant \p{W4-single} involves a single colour trace. One can also write a double-trace invariant, 
 \be
\lb{W4-double}
 S^{d=8}_2 \ \sim \int d\zeta^{-4} \varepsilon_{abcd}  \left\langle  W^{+a} W^{+b} \right\rangle 
\left\langle W^{+c} W^{+d} \right\rangle \, .
  \ee
 By including the hypermultiplet, both \p{W4-single} and \p{W4-double} can be completed to the $\cN = (1,1)$ on-shell
invariant actions. The explicit expressions can be found in Ref.\cite{BIS}. 

What is the physical relevance of these new invariants? As was explained earlier, 
they are not admissible
counterterms in the Wilsonean Lagrangian for the theory \p{L-11} --- there are no UV divergences at the 2-loop level.
However, the single-trace invariant {\it does} appear in the effective field theory actions for certain
string theories and, in particular, in the derivative expansion of the Born-Infeld effective action for 
coincident $D5$-branes \cite{Tseytlin}. 

\subsection{$d=10$ and beyond}

At the 3-loop level, logarithmic ultraviolet divergences and the associated counterterms do appear. 
In the pure gauge theory, one can write down two different structures, 
  \be
\lb{W4-10-single}
 S^{d=10}_1 \ \sim \ \left\langle \int dZ \, \varepsilon_{abcd} W^{+a} W^{-b} W^{+c} W^{-d} \right\rangle 
   \ee
and 
   \be
\lb{W4-10-double}
 S^{d=10}_2 \ \sim \ \int dZ \, \varepsilon_{abcd}  \left\langle  W^{+a} W^{-b}  \right\rangle 
\left\langle  W^{+c} W^{-d} \right\rangle \,. 
   \ee

($W^{-a} = \nabla^{--} W^{+a}$). 

In contrast to \p{W4-single} and \p{W4-double}, the actions  \p{W4-10-single} and 
\p{W4-10-double} are invariant
under the $\cN = (1,0)$ transformations off shell --- the integrands are not G-analytic, but generic  superfields 
of zero harmonic charge,
and the integral is done over the whole superspace. Incidentally, the extra $\theta^-$ integration brings
 about the extra two dimensions of mass, compared to \p{W4-single} and \p{W4-double}, so that 
the invariants \p{W4-10-single} and \p{W4-10-double} carry dimension
10 rather than 8.   Using a special harmonic 
{\it on-shell} superfield technique, suggested first
in \cite{BHS}, 
one has succeeded in expressing explicitly the on-shell $\cN = (1,1)$ supersymmetric completions
of the actions \p{W4-10-single} and  \p{W4-10-double} \cite{BIS}.    

And here we meet a puzzle. The existence of two different $d=10$ 
invariants suggests the presence of two different 
logarithmically divergent
3-loop contributions to the scattering amplitudes: a single-trace 
and a double-trace one. The explicit 
calculations of \cite{Bern-gauge} confirmed the absence of ultraviolet
 logarithms at the 1-loop and 2-loop levels, 
but did not confirm the presence of 
two different logarithmically divergent 3-loop structures --- only 
the single-trace structure was seen. We do not now 
understand why the single-trace counterterm is selected and the double-trace
 is not. Hopefully, a more meticulous study combining
the harmonic superspace methods with cohomology arguments of Ref.\cite{BHS} 
 could  provide an answer to this question. 

As was explained above, the full Wilsonean effective Lagrangian ${\cal L} = {\cal L}^{(d=4)} + 
{\cal L}^{(d=10)} + \ldots $ should be invariant under the modified supersymmetry transformations,
$\delta = \delta_0 + \delta_1 + \ldots $ , where $\delta_0$ was given in \p{trans-01}, and $\delta_1$ has
the order $\sim (g\partial)^6 \delta_0$, with $\partial$ meaning an extra 
spatial  derivative. The effective
Lagrangian can also include the terms of  still higher dimension, $\sim  {\cal L}^{(d=12)} + 
{\cal L}^{(d=14)} + \ldots $ The same concerns the modified transformations. The variations 
$\delta_0 {\cal L}^{(d=12)}$ and $\delta_0 {\cal L}^{(d=14)}$ should still vanish on mass shell. This follows
from the condition \p{onshell-01}, where one can include $ {\cal L}^{(d=12)}$ and $ {\cal L}^{(d=14)}$ 
into ${\cal L}_1$ and the corresponding higher-derivative terms in the supersymmetry transformations into
 $\delta_1$. The situation becomes more complicated at the level $d=16$.
 $ {\cal L}^{(d=16)}$ satisfies a more complicated condition \p{onshell-012} with nonzero 
$\delta_1 {\cal L}_1 \equiv \delta_1 {\cal L}^{(d=10)}$ and need not be on-shell supersymmetric. 
But the amplitudes {\it are} supersymmetric - cf. the discussion in Sect.~4.2.

 \section{Lessons for supergravity}
\setcounter{equation}0

As was mentioned, in Einstein's gravity, the first relevant
 counterterm \p{Lgrav-d6} 
has dimension $d=6$  and shows up at the 2-loop level. Since 40 years, 
it has been known that the structure \p{Lgrav-d6} cannot be supersymmetrized: 
the effective Lagrangian \p{Lgrav-d6} generates helicity-flip amplitudes, which
is not compatible with ${\cal N} = 1$ supersymmetry \cite{Grisaru}. 

At the 3-loop level, we can have the structure $\sim R^4 + \ldots$, which
is not protected by this argument \cite{Deser}. That means that 
logarithmic UV divergences may appear in the
 ${\cal N} = 1$ supergravity at the 3-loop level (though, 
to the best of our knowledge, this has not yet been directly checked).

Extended supersymmetries and, especially, the maximal $\cN = 8$ supersymmetry 
bring about further constraints. 
To establish them is
a more difficult task than for the extended $6D$ SYM theory 
discussed in Sect.~5,6. First, because the dimension of 
 the ``dangerous'' structures 
that one should study for supergravity is higher and the structures are 
more complicated. Second, because of a much wider ``gap'' between the superfield
description (which is possible only for the minimal $\cN = 1$ supersymmetry)
and the extended $\cN = 8$ on-shell supersymmetry, the presence or the absence
of which for the candidate counterterms one has to establish. 

Since 
\cite{Kallosh}, people have been aware of the presence of the counterterm of 
dimension $d=18$ that satisfies all the on-shell symmetries in the 
$\cN =8$ theory. It should bring about logarithmic 
UV divergences at eight loops. At that time, it was not clear, however,
 whether also some lower-dimensional counterterms 
(starting from the 3-loop level)
are or are not admissible.

The interest to this problem was resuscitated in the new century, 
owing to the 
works of  Bern's group that displayed the absence of logarithmic UV 
divergences at the 3-loop and then at the 4-loop level \cite{Bern-grav}. 
This fact needed to be explained. After several years of hard work (see e.g. 
\cite{Drummond,BHS,Beisert}),
people understood that the extended $\cN = 8$ supersymmetry and other related
on-shell symmetries exclude the presence of counterterms up to $d=14$. 
This means that on-shell scattering amplitudes should be free from logarithmic
UV divergences through 6 loops. 

The frontier of unknown is now pushed up to seven loops. An invariant of
dimension $d=16$, ${\cal L} \sim \partial^8 R^4 + \ldots$ which seems to 
satisfy all the on-shell symmetry requirements was constructed \cite{BHSV}. It
suggests the presence of UV logarithms at 7 loops.   

We cannot now be sure, however, that these divergences are indeed there, because there are at least 
three occasions where known theoretical arguments failed to justify certain noteworthy 
cancellations seen in ``experiment'' --- in explicit perturbative calculations. The first
such example was discussed in Sect.~6 --- the calculations displayed the absence of the divergences
associated with the   double-trace $d=10$ structure in 3-loop calculations in $6D$ SYM theory.
The second and the third examples are the extended, but not maximally extended    
$4D$ supergravities with $\cN = 4$  and $\cN = 5$. 
 For $\cN = 4$, one can build up a 3-loop on-shell invariant \cite{Deser} 
and, for $\cN = 5$ a 4-loop on-shell invariant \cite{Beisert,BHSV}. 
But the explicit calculations \cite{BDDH} displayed
the absence of the divergences at this level. 
 Another known example of unexpected cancellations refers to a certain
gauged supergravity where the beta function was shown to vanish at the 
1-loop level \cite{Duff}.

Maybe 7-loop divergences in the $\cN = 8$ supergravity 
are also killed
by an unknown yet reason, in which case the logarithmic UV divergences 
only appear starting from eight loops, as was anticipated
  back in 1980 \cite{Kallosh}...

Strictly speaking, and especially bearing in mind the just mentioned mismatch,
 indicating the absence of full understanding, we cannot be quite sure of that until explicit
7-loop and 8-loop calculations in $\cN= 8$ supergravity are done. Unfortunately, there is a little hope
to see such calculations in the foreseeable future. Our bet, however, is that the logarithmic divergences appear at some
level, counterterms start playing a role and there is an infinite number of them. This means that 
$\cN = 8$ supergravity has essentially the same ultraviolet behaviour as Fermi theory and as chiral theory
and cannot be considered as  fundamental.

The question of whether higher-dimensional counterterms are relevant or not
is important. Indeed,
it is clear from the discussion in Sect.~2 that non-renormalizability represents a problem not
so much due to logarithmic or power UV divergences (they can in principle be removed 
order by order from physical observables), but due to an uncontrollable energy growth of amplitudes 
and cross-sections and impossibility to make perturbative calculations beyond the unitarity barrier.
 Such calculations are definitely impossible in the presence
of higher-dimensional counterterms: high-energy amplitudes and high-energy cross
sections would essentially depend in this case on an 
infinite number of dimensionful coupling constants.

 Were such extra constants  absent, one could contemplate a scenario where the theory would be 
defined
perturbatively at all energies.  
Suppose  (we know  that it is wrong, but suppose) 
that Fermi theory would not involve extra counterterms bringing about logarithmic UV divergences. 
 In this case, everything
depended on the physical renormalized Fermi's constant $G_F^{\rm phys}$. Consider
 the total cross section of $\nu e$
scattering. At the tree level, it grows with energy as 
 $\sigma_{\nu e} \sim G_F^2 s$.  
With loop corrections taken into account and 
renormalization  performed,  it would acquire the form
\be
\lb{sech-Fermi}
\sigma_{\nu e} \ =\  G_F^2 s \left( a_0 + a_1 G_F s + \ldots \right) \, .
  \ee

One could then speculate that the series in the R.H.S. of \p{sech-Fermi} 
might converge to a function 
$f(G_F s)$ that falled down with $s$, so that the global cross section 
did not grow as a power and the 
Froissart theorem were fulfilled.

One can ask whether something similar could happen in gravity, at 
least for first six 
loops where 
there are no counterterms.
 Unfortunately, gravity involves massless particles, this 
invalidates the
Froissart theorem and the whole reasoning above. The {\it total} 
graviton-graviton 
cross section is 
simply 
infinite by the same reason as the total Coulomb
 cross section is infinite --- due to the massless propagators
in the $t$-channel. One can calculate differential cross sections 
allowing for the 
creation of extra soft
gravitons, but it is a complicated object depending on several
 kinematic invariants 
including  $\omega_{\rm max}$ (the maximal allowed total energy for 
extra soft particles produced) 
\cite{Weinberg}.
We do not know how to define it in higher loops, if one wants to keep track
 not only of the infrared $\omega_{\rm max}$ - dependent part, but also
of the finite part.

Thus, the Planck mass barier is difficult (impossible?)
 to cross in $\cN = 8$ supergravity even in a hypothetical unprobable
case where for some reason all the coefficients of higher-dimensional 
counterterms vanish...


\section*{Acknowledgements}
I am indebted to  Z. Bern, G. Bossard, I. Buchbinder, J. Donoghue, M. Duff, 
S. Fedoruk, E. Ivanov and U. Lindstr\"om  for  useful comments.

\end{document}